\documentclass[pre,showpacs,amsmath,amssymb,floatfix]{revtex4}
\usepackage[english]{babel}
\usepackage{t1enc}
\usepackage{hyperref}
\usepackage{graphicx}
\def\p{\textrm{\bf p}}
\def\q{\textrm{\bf q}}
\def\E{{\cal E}}
\def\G{A}

\def\O{{\cal O}}
\def\H{{\cal B}}
\def\be{\begin{equation}}
\def\ee{\end{equation}}

\begin{document}
\title{An additive variant of Tsallis generalized entropy}
\author{M. P. Almeida} \email{murilo@fisica.ufc.br}
\affiliation{
Departamento de F\'{\i}sica, Universidade Federal do Cear\'{a}, Caixa
Postal 6030, 60455-900, Fortaleza, Cear\'{a}, Brazil }

\begin{abstract} This is an analysis of the additivity of the
entropy of thermodynamical systems with finite heat baths.  It is
presented an expression for the physical entropy of weakly interacting
ergodic systems, and it is shown that it is valid for both the
microcanonical (constant energy), the Boltzmann-Gibbs canonical
(infinite heat bath) and the Tsallis (finite heat bath) ensembles.
This physical entropy may be written as a variant of Tsallis entropy.
It becomes an additive function after a suitable choice of
additive constants, in a procedure reminiscent to the solution
presented by Gibbs to the paradox bearing his name.
\end{abstract}
\maketitle

\section{Introduction}

The non-extensivity is one of the major features of Tsallis entropy
and perhaps the reason for most of the criticism it
receives.
This property was recognized in the very
first paper \cite{tsallis88} where the generalized entropy has been proposed
in its discrete form
\begin{equation}
S_q(X)= \frac{1- \sum_{i=1}^N[p(X=x_i)]^q}{q-1},
\end{equation}
where $q$ is a real constant and $ p(X=x_i)$ is the probability
distribution of a discrete random variable $X$ that takes value on
$\{x_1,\ldots,x_N\}$.  
Tsallis \cite{tsallis88} obtained the {\em
non-extensive} (non-additive) relation
\begin{equation}\label{non-additivity}
S_q(X,Y)= S_q(X)+S_q(Y)-(q-1)S_q(X)S_q(Y)
\end{equation}
computing the entropy of the joint distribution of two
independently distributed random variables $X$ and $Y$.
Since then, the words ``extensive'' and ``additive'' have been loosely
used as synonymous in the literature about Tsallis statistics
\cite{touchette}. However, following Touchette's \cite{touchette}
guidelines for the usage of these terms, the property expressed by
Eq.~(\ref{non-additivity}) is called  here ``non-additivity''.

In a recent paper, Abe \cite{PhysA305_62} has shown
that, for composite systems resulting from the joint observation of
two independently distributed sub-systems, the "physical entropy" is
additive at the macroscopic level, even in the context of Tsallis
statistics, where by ``physical entropy'' he means the function (Rényi
entropy)
\begin{equation}
S^* = \frac{\ln(1+(1-q)S_q)}{1-q}=\frac{\ln(\sum_{i=1}^N[p(X=x_i)]^q)}{1-q},
\end{equation}
where $S_q$ is the Tsallis entropy.
 
Power-law (Tsallis) canonical distributions have been connected to the
finitude of the dimension of the phase space of the heat bath in
\cite{Plastino_94}, \cite{Alm01} and \cite{Adib}.

The issue of additivity within Tsallis statistics has been addressed
in the scientific literature up to now assuming {\em the joint
observation of independent random variables} as done in the derivation
of Eq.~(\ref{non-additivity}).  Regardless of the validity of this
rule for non-interacting systems and for ``weakly interacting
systems'' in the thermodynamic limit, the same can not be generally
applied for finite thermodynamic systems.  We say that two systems are
in weak interaction when the Hamiltonian of the composite system is
just the sum of the Hamiltonians of its sub-components,
$H_{1+2}=H_1+H_2$. As a consequence, the internal energy $(U=\int H
\rho d\q d\p)$ of such a system in an equilibrium state $\rho$ is an
additive variable.

Consider, for instance, the observation of the 
energy.
When one joins (through weak interaction) two independent sub-systems,
$A_1$ and $A_2$, the distribution of the energy of the composite
system $A$, ($E_A=E_{A_1} + E_{A_2}$), is given by the convolution of
the sub-component's distributions, $\rho(E_A)=\rho(E_{A_1})\otimes
\rho(E_{A_2})$, which is generally different from the distribution
of the joint observation of $E_{A_1}$ and $E_{A_2}$,
$\rho(E_{A_1},E_{A_2})=\rho(E_{A_1})\rho(E_{A_2})$.  In other words,
when observing $E_A$, it is not necessary to know how much of it comes
from system $A_1$ or $A_2$, but only that the sum of these
contributions adds up to the observed quantity. In addition to this,
even the range of the energy value of each sub-system changes upon the
composition: If each separate sub-system initially has
internal energy equal to $E_0$, after their merging the composite
system will have energy equal to $2E_0$ and the energy of each
sub-system will now fluctuate in the interval $[0, 2E_0]$. Therefore,
the joint distribution of the simultaneous observation of two
independent systems $A_1$ and $A_2$ does not represent the composite
thermodynamic system $A_1+A_2$.  Since the distribution of a sum of
independent random variables is a ``convolution'' of the distributions
of the component variables, the additivity of the energy implies that
the correct distribution for the composite system shall be obtained
through a convolution operation.  Only in the thermodynamic limit,
when the sub-systems become infinite, the resulting distribution
will factor into the product of the component distributions. {Moreover, it
can be rigorously proof that for systems in weak interaction the
Boltzmann-Gibbs canonical distribution (exponential) is the only one
for which the convolution of the components equals their product}.

We will derive a {\em physical entropy function} $S$ that is additive
at the equilibrium distribution even for finite systems. For a given
system $A$ resulting from a {\em weak combination} of two sub-systems
$A_1$ and $A_2$, (all of them in equilibrium), the entropy function
$S$ satisfies the {\em additive} rule $S(A)=S(A_1)+S(A_2)$. We will
explicitly evaluate $S$ for canonical ergodic systems.  This analysis
will be based on the canonical ensemble theory of ergodic systems as
presented by Kinchin~\cite{Kinchin}, whose basic elements we recollect
in Section {\bf II}.  The present work is a follow up of the study
started in~\cite{Alm01} where it was presented a one-to-one connection
between the canonical distribution and the form of the entropy, and
the Tsallis entropy was derived from first principles. This
alternative approach for Tsallis statistics has been already used by
Andrade et al.~\cite{PRE65_036121} to formulate a molecular dynamics
scheme for Hamiltonian systems exhibiting power-law distributions,
generalizing the extended Hamiltonian approach presented by
Nosé~\cite{Nose}. This setup was also used by Adib et al.~\cite{Adib}
to analyse the statistical behavior of systems with homogenous
Hamiltonians.

\section{Ergodic systems}

For a generic system with Hamiltonian $H=H(\q,\p)$ and phase space
$\Xi$, let $V(E)$ denote the volume of the region of $\Xi$ where
$H(\q,\p) \le E$, $\Omega(E)= \partial V(E)/\partial E$ be the {\em
structure function} \cite{Kinchin}, and $\beta(E)$ be the function
\begin{equation}\label{beta}
\beta(E)=\frac{d}{dE}\ln\Omega(E).
\end{equation}

The distribution on the surface of constant energy ($\Sigma(E)$) that is
invariant under the motion of the system is called {\em
microcanonical} and has density given by (see Chapter {\bf II} of
\cite{Kinchin})
\be\label{mc_rho}
\rho^\textrm{(mc)}(\q, \p)=\frac{1}{\Omega(E) ||\nabla H(\q,\p) ||},
\ee
where $||\nabla H(\q,\p)||$ is the norm of the gradient of the Hamiltonian
$H(\q,\p)$.

For a system $\G$ which is the {\em weak combination} of two
sub-systems $\G_1$ and $\G_2$ (denoted by $\G=\G_1\uplus\G_2$), we
have that $H(\q,\p)= H_1(\q_1,\p_1) + H_2(\q_2,\p_2)$ with no
interaction term between $\G_1$ and $\G_2$ in the Hamiltonian, i.e., 
the energies of the sub-systems are additive. In that case the
canonical distribution in $\Xi_1$, when the total energy of the
composite system $\G$ is equal to $E$, is given by the density
function
\cite{Kinchin}
\begin{equation} \label{rho_1}
\rho_1(\q,\p)=\frac{\Omega_2(E-H_1(\q,\p))}{\Omega(E)}.
\end{equation}
The structure function of $\G$ satisfies the convolution relation
\begin{equation}\label{convolution}
\Omega(E)=\int_0^E \Omega_1(\xi)\Omega_2(E-\xi) d\xi.
\end{equation}
Observe that the microcanonical distribution on $\Sigma_1(E_1)$ is the
conditional distribution of $\rho_1$ obtained by fixing the value of
$H_1(\q,\p)=E_1$, which results in
\be\label{mc_rho_1}
\rho_1^\textrm{(mc)}(\q,\p)=\frac{1}{\Omega_1(E_1) ||\nabla H_1(\q,\p) ||}.
\ee
It has been shown in \cite{Alm01} that the form of the canonical
$\rho_1$ depends on the value of the parameter $q_2$, which is related
to $\beta_2(E)$ through the expression
\begin{equation}\label{dbeta}
\frac{d}{dE}\left(\frac{1}{\beta_2(E)}\right)=q_2-1.
\end{equation}
When $q_2=1$, $\rho_1$ is an exponential
(Boltzmann-Gibbs) and, when $q_2\ne 1$, $\rho_1$ is a power-law
(Tsallis) distribution. Identifying system $\G_2$ with the heat bath
$\H$ and integrating Eq.~(\ref{dbeta}) with the initial condition
$\Omega_2(0)=0$ we get that
\begin{equation}\label{Vi}
\Omega_2(E)=C_0 E^{1/(q_2-1)},
\end{equation}
where $C_0$ is a constant, and consequently
\begin{equation}\label{betai}
\beta_2(E)=\frac{1}{(q_2-1)E}.
\end{equation} 

We are going to consider that the elementary systems appearing from
now on have structure functions in the form $\Omega(E) = C E^{1/(q-1)}$, with
specific constants $C$ and $q$ for each one. The parameter $q$,
which is a signature of the system's phase space geometry, may vary
from system to system.  Indeed, Almeida et al.~\cite{temperature} have
showed that the weak composition of two systems $\O$ and $\H$,
with parameters $q_{\O}$ and $q_{\H}$, respectively, results in a system
$\G$ with parameter $q_\G$ given by
\begin{equation}\label{q_relation1}
\frac{q_\G}{q_\G-1}=\frac{q_{\O}}{q_{\O}-1}+ \frac{q_{\H}}{q_{\H}-1},
\end{equation}
and that the temperature definition obeying the theorem of
equipartition of energy leads to the expressions
\begin{eqnarray}
kT &=& \left(\frac{q_\G-1}{q_\G}\right) E_\G \label{phys-temp}\\ 
   &=& \left(\frac{q_{\O}-1}{q_{\O}}\right) <H_{\O}> \label{phys-temp1}\\
   &=& \left(\frac{q_{\H}-1}{q_{\H}}\right) <H_{\H}>. \label{phys-temp2}
\end{eqnarray} 

Let's consider a weak combination of independent canonical sub-systems
$A_i$, each one having an observable sub-system $\O_i=\O(A_i)$ and a heat
bath $\H_i=\H(A_i)$.  The observable and the heat bath of the composite
system $A$ are, respectively, the weak combinations of the observables
and of the heat baths of the sub-systems, i.e.,
$A=\uplus_{i} A_i$, $\O(A)=\uplus_{i} \O(A_i)$, and $\H(A)
=\uplus_{i}\H(A_i)$.  Identifying the observable part $\O(=\O(A))$ and
the heat bath $\H(=\H(A))$, respectively, with the systems $\G_1$ and
$\G_2$ in Eq.~(\ref{rho_1}), we write the canonical distribution of
$A$ as
\begin{equation} \label{rho_oa}
\rho_{\O}(H_{\O})=\frac{\Omega_{\H}(E_A-H_{\O})}{\Omega_A(E_A)},
\end{equation}
where $E_A$ is the total energy of $A$ , $H_{\O}$ is the energy of
its observable sub-system, and $\Omega_A$ and $\Omega_{\H}$ are the
structure functions of the total system $A$ and of its heat bath
$\H$, respectively.
Writing $\rho_{\O}$ in the usual Tsallis density form 
\be \label{rho_o}
\rho(H_{\O})=\rho_{\O}(H_{\O})= \rho_0 \left[ 1 - (q_\H -1) \beta H_\O\right]^{1/(q_\H-1)},
\ee
we have that
\begin{equation}\label{rho_o0}
\rho_0 = \frac {\Omega_\H(E_\G)} {\Omega_\G(E_\G)} 
\end{equation}
and
\begin{equation}\label{qbeta}
\beta=\beta_\H(E_\G) = \frac{1}{ (q_\H-1)E_\G}.
\end{equation}
Moreover, the equivalent to
Eq.~(\ref{q_relation1}) reads as
\begin{equation}\label{q_relation2}
\frac {q_A} {q_A-1}=\frac {q_{\O}} {q_{\O}-1}+ \frac {q_{\H}} {q_{\H}-1},
\end{equation}
where $q_A$, $q_{\O}$ and $q_{\H}$ are the parameters $q$ of $A$,
$\O$ and $\H$, respectively. 

Applying Eq.~(\ref{q_relation1}) recursively to $A$, $\O$ and $\H$ we get the relations
\begin{equation}\label{q_combination}
\frac {q_A} {q_A-1}=\sum_i \frac {q_{A_i}} {q_{A_i}-1}
\end{equation}
\begin{equation}
\frac {q_{\O}} {q_{\O}-1}=\sum_i \frac {q_{\O_i}} {q_{\O_i}-1}\label{qo_combination}
\end{equation}
and
\begin{equation}\label{qb_combination}
\frac {q_{\H}} {q_{\H}-1}=\sum_i \frac {q_{\H_i}} {q_{\H_i}-1}.
\end{equation}

\section{Physical entropy}
For a system with Hamiltonian $H$ and probability density function
$\rho$ let $U$ be the internal energy, $W$ be the work of the external
forces and $Q$ be the heat supplied to the system. In terms of the
microscopic functions we have that
\be  \label{U}
U=\int_{\Xi} H\rho d\q d\p.
\ee
The first law of thermodynamics states that
\be
\delta U=\delta W + \delta Q.
\ee
Computing the variation $\delta U$ from its microscopic definition, Eq.~(\ref{U}),
we get that
\be\label{deltaU}
\delta U  =  \int_{\Xi} (\delta H) \rho d\q d\p + \int_{\Xi} H (\delta \rho) d\q d\p. 
\ee
The first term on the right hand side of this equation is the variation of the work $\delta W$ and so we get that
\be\label{deltaQ}
\delta Q = \int_{\Xi} H (\delta \rho) d\q d\p.
\ee
The second law of thermodynamics states that there are an exact differential functions $S$, called entropies, such
that 
\be\label{deltaS}
\delta S = \frac{1}{T} \delta Q,
\ee
where $T$, the integrating factor, is a temperature scale.

As pointed out in a previous work \cite{Alm01}, the functional form of
the physical entropy is determined by the canonical distribution:
Given a canonical density function in $\Xi$, $\rho=f(H)$, with a
non-negative monotonic function $f(x)$ defined for $x\ge 0$, the
functions (entropies) that satisfy the second law of thermodynamics
are given by
\begin{equation}\label{S-formula}
S(\rho)=\int_{\Xi} F(\rho) d\q d\p
\end{equation}
with
\begin{equation}\label{F-formula}
F(\rho) = \frac {1} {T} \int_0^\rho f^{-1}(\xi)d\xi + \psi \rho,
\end{equation} 
where $\psi$ is an arbitrary constant. As will be seen later, it is
convenient to choose $\psi=-k$, where $k$ is the Boltzmann constant.

Applying the above procedure to the Tsallis density function 
\begin{equation}\label{T-rho}
\rho(H) = \rho_0[1-(q_\H-1)\beta H]^{1/(q_\H-1)},
\end{equation}
where $(q_\H-1)\beta=1/E$ is the inverse of the total energy of the system, considering that 
\begin{equation}\label{rho_0}
\rho_0=\rho(0)=\left\{\int_\Xi [1-(q_\H-1)\beta H]^{1/(q_\H-1)}d\q d\p\right\}^{-1},
\end{equation}
is constant, we get the function
\begin{equation}\label{TS-generic1}
\E(\rho)= \frac{1}{T\ (q_\H-1)\beta}\left[1-\frac{\rho_0} {q_\H}
\int_\Xi\left(\frac{\rho}{\rho_0}\right)^{q_\H} d\q d\p\right]- k,
\end{equation} 
which is not the entropy yet, because of the imposed condition that $\rho_0$ be constant.
Collecting and rearranging terms we get the alternative expression
\begin{eqnarray}
\E(\rho)=& \frac{1}{q_\H T\beta}\left\{\frac{\rho_0} {q_\H-1}
\int_\Xi\left[\left(\frac{\rho}{\rho_0}\right)-\left(\frac{\rho}{\rho_0}\right)^{q_\H} \right] d\q d\p\right\} \nonumber\\
 &+\left(\frac{1}{q_\H T\beta}- k\right).\label{E}
\end{eqnarray}
This expression  may be written also as
\begin{equation}\label{E-3}
\E(\rho)= \left(\frac{1}{q_\H T\beta}\right)\ \rho_0\  S_{\textrm{TS}}\left(\rho/\rho_0 \right)
+\left(\frac{1}{q_\H T\beta}- k\right),
\end{equation}
where
\begin{equation}\label{S-TS}
 S_{\textrm{TS}}(\rho/\rho_0)=\frac{1} {q_\H-1}
\int_\Xi\left[\left(\frac{\rho}{\rho_0}\right)-\left(\frac{\rho}{\rho_0}\right)^{q_\H} \right] d\q d\p
\end{equation}
is the Tsallis entropy evaluated at the ratio $\rho/\rho_0$. Observe
in Eq.~(\ref{E-3}) the splitting of the roles played
by the density function form, $\rho/\rho_0$, and by the
normalization value $\rho_0$.

Now, considering that both $\rho$ and $\rho_0$ vary along the process,
the evaluation of $\delta\E$ will produce the sought expression for
the entropy. From Eq.~(\ref{TS-generic1}) we have that
\begin{eqnarray}
\delta \E = & \frac{1}{T\ (q_\H-1)\beta}\left[-\int_\Xi\left(\frac{\rho}{\rho_0}\right)^{q_\H-1}(\delta\rho)  d\q d\p\right. +\nonumber \\
& \left.(\delta \rho_0)\frac{q_\H-1}{q_\H}\int_\Xi\left(\frac{\rho}{\rho_0}\right)^{q_\H}d\q d\p\right].
\end{eqnarray}
Substituting the expressions for $\rho$ and $\rho_0$, and considering the temperature expressions Eqs.~(\ref{phys-temp})-(\ref{phys-temp2})
we get that 
\be
\delta \E =  \frac{1}{T}\int_\Xi H (\delta\rho)  d\q d\p + k \delta(\ln\rho_0) =\frac{1}{T} \delta Q + k \delta(\ln\rho_0),
\ee
and from it we conclude that
\be
\delta S = \frac{1}{T} \delta Q = \delta \E - k \delta(\ln\rho_0),
\ee
and so the generic entropies  are  
\be\label{entropia}
S(\rho) = \E(\rho) - k \ln\rho_0 + C,
\ee
where $C$ is an arbitrary constant.

For a fixed value of $D=q_\H T\beta$, the maximization of
$S(\rho)$ under the conditions 
\begin{eqnarray}
&\int\rho d\q d\p =1, \\
&\int H\rho d\q d\p =U,
\end{eqnarray}
and
\be
\rho_0=\rho(0),
\ee
results in the Tsallis density function $\rho(H)$ given by Eqs.~(\ref{T-rho}) and (\ref{rho_0}).

\subsection{Additivity of $\E$}
Let's compute the explicit value of $\E_\G(\rho)$ for a system
$\G(=\O\uplus \H)$ whose canonical distribution on $\Xi_\O$,
$\rho(H_\O)$, is given in Eq.~(\ref{rho_o}).
From a direct integration we get 
\begin{equation}\label{part1}
\rho_0\int_{\Xi_\O} \left(\frac{\rho}{\rho_0}\right)^{q_\H}d\q d\p =
\left[ 1 -\frac{<H_\O>}{E_\G}\right],
\end{equation} 
where we adopt the notation $<f>=\int_{\Xi_\O} f \rho d\q d\p$ for
the average of a generic function $f$. 

From Eqs.~(\ref{qbeta}) and (\ref{phys-temp}) we get that 
\begin{equation}\label{part2}
\frac{1}{T\ (q_\H-1)\beta} = k \left(\frac{q_\G}{q_\G-1}\right).
\end{equation}
Substituting Eqs.~(\ref{part1}) and (\ref{part2}) in
Eq.~(\ref{TS-generic1}) and using Eq.~(\ref{q_relation2}) one gets
that
\begin{equation} \label{E-explicit}
\E_\G(\rho)= k \left(\frac{q_\O} {q_\O-1}\right).
\end{equation} 
Using the combination rule for the $q_\O$'s,
Eq.~(\ref{qo_combination}), we observe that under {\em weak
combination} of sub-systems
the function $\E_\G$ is additive. In general, if $\G$ is a
(weak) combination of $N$ sub-systems $\G_i$, $i=1,\ldots,N$,
$(\G=\G_1\uplus \G_2 \uplus\ldots \uplus\G_N)$, all at the same
temperature $T$, then
\begin{equation} \label{entropy_additivity}
\E_\G(\rho) = \sum_{i=1}^N \E_{\G_i}(\rho_i).
\end{equation} 

Notice that $\E_\G(\rho)$ does not depend on the parameter $q_\H$,
hence Eq.~(\ref{E-explicit}) is valid in the limit case
$\q_\H\to 1$ (Boltzmann-Gibbs), which can be also verified through a
direct evaluation of Eq.~(\ref{E-bg}) with
$\rho=\rho_0\exp(-\beta H)$.  Observe that Eq.~(\ref{phys-temp1})
together with Eq.~(\ref{E-explicit}) leads to the relation
between $\E_\G$, the internal energy $U=< H_\O >$ and the
temperature $T$,
\begin{equation}
\E_\G(\rho)=\frac { U } {T}.
\end{equation}

Substituting the alternative forms of $\E(\rho)$ into Eq.~(\ref{entropia}) 
we obtain
\be\label{entropia1a}
S(\rho) = k\left(\frac{q_\O}{q_\O-1}\right) - k \ln\rho_0 + C
\ee
and
\be\label{entropia1}
S(\rho) = \frac{U}{T} - k \ln\rho_0 + C.
\ee
This last expression  was also obtained  through a
different route by Potiguar and Costa \cite{fabricio02a}. 

Notice that the constants $q_\H$ and $\beta$ appearing in
Eq.~(\ref{E}) are associated with the heat bath and that
\begin{eqnarray}
\frac{1}{q_\H T\beta}& = & k\left(\frac{q_\G}{q_\G-1}\right) \left(\frac{q_\H-1}{q_\H}\right) \\
& = & k\left(\frac{q_\O}{q_\O-1}\right) \left(\frac{q_\H-1}{q_\H}\right) + k \label{qkt}
\end{eqnarray}
is determined by the constants $q_\O$ and $q_\H$.  From this last
equation, we see that the commonly used relation $kT\beta=1$ is valid
only in the limit $q_\H\to 1$, i.e., in the Boltzmann-Gibbs
statistics.  This fact has been overseen by some authors who
inadvertently adopted $kT=1/\beta$ within Tsallis statistics, which
might have led to misleading results.

\subsection{Microcanonical ensemble}
The microcanonical ensemble is represented by a constant energy
$H_\O$ and its density function in this surface $\Sigma_\O$  is
\be\label{mcdensity}
\rho^\textrm{(mc)}(H_\O)=\frac{1}{\Omega_\O(H_\O) ||\nabla H_\O ||}.
\ee
Hence, we may consider that in Eq.~(\ref{mcdensity})  $\rho_0=(\Omega_\O)^{-1}$  and obtain from
Eq.~(\ref{entropia1a}) that    
\be\label{entropiamc}
S(\rho^\textrm{(mc)}) = k \ln\left[\Omega_\O(H_\O)\right]
+ \left[k\left(\frac{q_\O}{q_\O-1}\right) + C\right],
\ee
which is the Boltzmann entropy for the microcanonical ensemble plus constants.

\subsection{Boltzmann-Gibbs canonical ensemble}
The Boltzmann-Gibbs canonical ensemble considers an infinite heat bath, which
is represented by the limit condition $\q_\H\to 1$ keeping $\beta$ fixed.
Taking this limit in Eq.~(\ref{rho_o}) and  in either of the 
Eqs.~(\ref{TS-generic1}) or (\ref{E}) we arrive at
\be\label{BGcanonical}
\rho(H_\O)=\rho_0\exp(-\beta H_\O)
\ee
and
\begin{equation}\label{E-bg}
\E(\rho)=\frac{1}{T\beta} \left\{ 1- \int_\Xi \rho \ln(\rho) d\q d\p +\ln(\rho_0) \right\} - k,
\end{equation}
from which, considering also that in the limit $q_\H\to 1$ we have 
$1/(T\beta)=k$ (see Eq.~(\ref{qkt}) below),
we get that 
\be
S=\E(\rho)-k\ln(\rho_0)+C=-k\int_\Xi \rho \ln(\rho) d\q d\p +C
\ee
 is the Boltzmann-Gibbs-Shannon entropy plus an arbitrary constant $C$. 

\subsection{Additivity of $k\ln\rho_0$}
Let's consider that the system $\G$ is the weak combination of $n$
identical sub-systems $\G_i$, $i=1,\ldots,n$, which by their turn are
the weak combination of an observable $\O_i$ and a heat bath $\H_i$.
Let $q$, $q_1$ and $q_2$ be the $q$-parameters of the $\G_i$, $\O_i$
and $\H_i$, respectively. Therefore, we have that the parameters for
the system $\G$, its observable part $\O$, and its heat bath $\H$ are
respectively $q_\G$, $q_\O$ and $q_\H$, and they satisfy the relations
of Eqs.~(\ref{q_combination})-(\ref{qb_combination}), i.e.,
\begin{equation}\label{q_combination1}
\frac {q_A} {q_A-1}=\sum_i\frac {q_{A_i}} {q_{A_i}-1}=n \frac {q} {q-1}
\end{equation}
\begin{equation}
\frac {q_{\O}} {q_{\O}-1}=\sum_i \frac {q_{\O_i}} {q_{\O_i}-1}=n \frac {q_1} {q_1-1} \label{qo_combination1}
\end{equation}
and
\begin{equation}\label{qb_combination1}
\frac {q_{\H}} {q_{\H}-1}=\sum_i \frac {q_{\H_i}} {q_{\H_i}-1}=n \frac {q_2} {q_2-1}.
\end{equation}

Computing the structure function of the system $\G$ using Eq.~(\ref{convolution}) we obtain
\be\label{omegaa}
\Omega_\G(E_\G)=\Omega_\O(E_\G)\Omega_\H(E_\G)E_\G \frac{\Gamma\left(\frac{q_\O}{q_\O-1}\right)\Gamma\left(\frac{q_\H}{q_\H-1}\right)}
{\Gamma\left(\frac{q_\G}{q_\G-1}\right)}
\ee
Therefore, 
\be\label{rho01}
\rho_0= \frac{\Omega_\H(E_\G)}{\Omega_\G(E_\G)}= {}\frac{\Gamma\left(\frac{q_\G}{q_\G-1}\right)}{\Omega_\O(E_\G) E_\G\Gamma\left(\frac{q_\O}{q_\O-1}\right)\Gamma\left(\frac{q_\H}{q_\H-1}\right)}
\ee
Recursively applying  Eq.~(\ref{convolution}) to compute $\Omega_\O(E_\G)$ we get that
\be\label{omegao}
\Omega_\O(E_\G)E_\G{\Gamma\left(\frac{q_\O}{q_\O-1}\right)}= \left[C_0\Gamma\left(\frac{q_1}{q_1-1}\right) E_\G^{\frac{q_1}{q_1-1}}
\right]^n,
\ee
where $C_0$ is the multiplicative constant appearing in the structure functions $\Omega_{\O_i}$.
Plugging (\ref{omegao}) into (\ref{rho01}) we obtain
\be\label{rho02}
\rho_0= \frac{\Gamma\left(\frac{q_\G}{q_\G-1}\right)}{\Gamma\left(\frac{q_\H}{q_\H-1}\right)}\left[C_0\Gamma\left(\frac{q_1}{q_1-1}\right) E_\G^{\frac{q_1}{q_1-1}}
\right]^{-n}.
\ee 
>From Eq.~(\ref{phys-temp}) the total energy $E_\G$ in terms of the temperature and $q$ is
\be\label{Ea}
E_\G=k T \left(\frac{q_\G}{q_\G-1}\right)= nkT\left(\frac{q}{q-1}\right),
\ee
>From these two last equations we get that
\begin{eqnarray}
-k \ln \rho_0& = & k\ln \left[\frac{\Gamma\left(\frac{n q_2}{q_2-1}\right)}
{\Gamma\left(\frac{n q}{q-1}\right)}\right]\label{log1} \\
& & + k n \left(\frac{q_1}{q_1-1}\right) \ln[n]\label{log2} \\
& & + k n \left(\frac{q_1}{q_1-1}\right) \ln \left[kT\left(\frac{q}{q-1}\right)\right]\label{log3}\\
& & + k n \ln\left[C_0 \Gamma\left(\frac{q_1}{q_1-1}\right)\right]. \label{log4}
\end{eqnarray}
The terms in Eqs.~(\ref{log3}) and (\ref{log4}) are additive since
they are linear in $n$, but the terms in (\ref{log1}) and (\ref{log2})
are non-additive. We can eliminate this non-additivity by taking the
constant $C$ appearing on Eq.~(\ref{entropia}) equal to the negative of the sum
of all terms on the right hand side of Eqs.~(\ref{log1})- (\ref{log4}) that depend, besides of $n$, only on the parameters
related to the smallest system sub-components, viz., $q$, $q_1$, $q_2$
and $C_0$.  This adequate choice of the constant $C$ is analogous to
the solution presented by Gibbs to the paradox bearing his name. In this way,
we are left solely with the only term depending on $T$,
\be\label{finalrho}
-k\ln\rho_0 +C= k n \left(\frac{q_1}{q_1-1}\right) \ln[T]= k\left(\frac{q_\O}{q_\O-1}\right)\ln[T].
\ee\label{entropia2}
Combining Eqs.~(\ref{entropia}), (\ref{E-explicit}) and (\ref{finalrho}) the resulting entropy function becomes
\be
S= k\left(\frac{q_\O}{q_\O-1}\right)\left[1+\ln[T]\right]=\frac{U}{T}\left[1+\ln[T]\right].
\ee
Since the temperature $T$ is an intensive variable and the internal
energy is additive, this entropy form is additive.

\section{Conclusions}
We have seen that both the microcanonical and the classical canonical
(Boltzmann-Gibbs) distributions can be obtained from the Tsallis
distribution. We have obtained also an expression for the ``physical
entropy'' of weak-interaction systems with finite heat baths ($q_\H\ne
1$) and showed that, for systems with constant energy, it yields the
classical microcanonical Boltzmann entropy and that when the heat bath
gets infinite ($q_H\to 1$), it becomes the classical
Boltzmann-Gibbs-Shannon entropy.  Moreover, we saw that this
``physical entropy'' may be written as a variant of Tsallis entropy
plus the product of the Boltzmann constant and the logarithm of the
partition function. In addition to this, we saw that it may be turned
into an additive variable by adding appropriate constants in a
procedure reminiscent to what was done by Gibbs to overcome the Gibbs
paradox of the microcanonical entropy for ideal gases.

Our results strength the physical meaning of the generalized entropy
by showing that the generalized thermostatistics fills the gap between
the microcanonical and the Boltzmann-Gibbs canonical ensembles. These
results are summarized in the following frame:
\begin{center}
\begin{tabular}{|c|c|c|c|}
\hline 
Ensemble & microcanonical & Tsallis & BG canonical \cr
\hline 
heat bath & --- & finite & infinite \cr
\hline
constraint & $H_\O=\textrm{constant}$  & $q_\H\ne 1 $ & $q_\H\to 1$ \cr
\hline 
density $\rho$ & $\rho_0[||\nabla H||]^{-1} $ & $\rho_0[1-(q_\H-1)\beta H_\O]^{1/(q_\H-1)}$ &
$\rho_0\exp(-\beta H_\O)$ \cr
\hline
Entropy & \multispan{2}{
\mbox{ $ S= k\left(\frac{q_\O}{q_\O-1}\right)-k\ln[\rho_0] $}} &  \cr 
\hline
\end{tabular}
\end{center}

The analysis presented here aimed at the additivity of the entropy
within thermodynamic systems and it was not intended to contest the
validity of Eq.~(\ref{non-additivity}), which is applicable for the
original Tsallis entropy $S_{\textrm{TS}}$ under joint observation of
independent random variables. It was also seen that when weakly
combining systems the parameter $q$ varies with the dimensionality of
the system's phase space following the additive rule for weak
interactions (Eq.~(\ref{q_relation1})).

\section{Acknowledgments}
This work was partially supported by CNPq (Brazil).


\begin{thebibliography}{99}

\bibitem{tsallis88} C. Tsallis, J. Stat. Phys. {\bf 52} (1988) 479 .

\bibitem{touchette}
H. Touchette, Physica A {\bf 305} (2002) 84.

\bibitem{PhysA305_62}
S. Abe, Physica A {\bf 305} (2002) 62.

\bibitem{Plastino_94} A. R. Plastino, A. Plastino, Phys. Lett. A, {\bf 193} (1994) 140.

\bibitem{Alm01}M. P. Almeida,
Physica A {\bf 300} (2001) 424.

\bibitem{Kinchin} A. I. Kinchin, {\it Mathematical Foundations of Statistical
Mechanics} (Dover, New York, 1949).

\bibitem{PRE65_036121}  J. S. Andrade, Jr., M. P. Almeida, A. A. Moreira, and G. A. Farias,
Phys. Rev. E {\bf 65} (2002) 036121.

\bibitem{Nose} S. Nosé, J. Chem. Phys. {\bf 81} (1984) 511.

\bibitem{Adib} Artur B. Adib, Andre A. Moreira, Jose S. Andrade Jr., Murilo P. Almeida,
Tsallis thermostatistics for finite systems: a Hamiltonian approach, cond-mat/0204034, (2002).

\bibitem{temperature} M. P. Almeida, F. Q. Potiguar, U. M. S. Costa,
Microscopic analog of temperature within nonextensive
thermostatistics, cond-mat/0206243 (2002).  

\bibitem{fabricio02a}
F. Q. Potiguar, U. M. S. Costa, Thermodynamics arising from Tsallis
thermostatistics, cond-mat/0208357 (2002).
  
\end{thebibliography}
\end{document}